\documentclass[useamsfonts]{pasj00}
\usepackage{graphicx}

\SetRunningHead{HASHIMOTO ET AL.}
	       {NEAR-INFRARED POLARIMETRY OMC-1S REGION}
	       
\begin{document}
		 
		 \title{Near-Infrared Polarization Images of\\ 
		   The Orion Molecular Cloud 1 South Region}
		 \author{Jun \textsc{HASHIMOTO},\altaffilmark{1,2} Motohide \textsc{TAMURA},\altaffilmark{1}
		   Ryo \textsc{KANDORI},\altaffilmark{1} Nobuhiko \textsc{KUSAKABE}\altaffilmark{1}}
		 \author{Yasushi \textsc{NAKAJIMA},\altaffilmark{1} Shuji \textsc{SATO},\altaffilmark{3} 
		   Chie \textsc{NAGASHIMA},\altaffilmark{3} Mikio \textsc{KURITA}\altaffilmark{3} }
		 \author{Tetsuya \textsc{NAGATA},\altaffilmark{4} Takahiro \textsc{NAGAYAMA},\altaffilmark{4} Jim \textsc{HOUGH}\altaffilmark{5}}
		  
		 \altaffiltext{1}{National Astronomical Observatory, 2-21-1 Osawa, Mitaka, Tokyo 181-8588; 
		   hashmtjn@optik.mtk.nao.ac.jp}
		 \altaffiltext{2}{Department of Physics, Tokyo University of Science, 1-3, Kagurazaka, Sinjuku-ku, Tokyo 162-8601}
		 \altaffiltext{3}{Department of Astrophysics, Nagoya University, Chikusa-ku, Nagoya 464-8602}
		 \altaffiltext{4}{Department of Astronomy, Kyoto University, Sakyo-ku, Kyoto 606-8502}
		 \altaffiltext{5}{Centre for Astrophysics Research, University of Hertfordshire, Hatfield HERTS AL10 9AB, UK}

		 \maketitle
		 \begin{abstract}
		   We present the polarization images in the $J$, $H$, \& $Ks$ bands of the Orion Molecular Cloud 1 South
		   region. The polarization images 
		   clearly show at least six infrared reflection nebulae (IRNe) which are barely seen 
		   or invisible in the intensity images. Our polarization vector images also identify  
		   the illuminating sources of the nebulae: IRN 1 \& 2, IRN 3, 4, \& 5, and IRN 6
		   are illuminated by three IR sources, Source 144-351, Source 145-356, and Source 136-355, respectively. 
		   Moreover, our polarization images suggest the candidate driving sources of the optical Herbig-Haro objects 
		   for the first time;
		   HH529, a pair of HH202 and HH528 or HH 203/204, HH 530 and HH269 are 
		   originated from Source 144-351, Source 145-356, and Source 136-355, respectively. 
		 \end{abstract}

\section{Introduction}
The Orion nebula region has two luminous star-forming cloud cores which
lie immediately behind the young (1 Myr) Orion Nebula Cluster (ONC) centered 
on the bright Trapezium OB stars. One is the $\sim 10^{5} L_{\odot}$ BN/KL 
region in Orion Molecular Cloud 1 (OMC-1)
and the other is the $\sim 10^{4} L_{\odot}$ OMC-1S region.
The BN/KL region has a number of compact infrared sources (IRc), of which only BN, IRc2, 
and IRc9 are believed to be self luminous, and are associated with a wide-angle 
bipolar outflow, powerful masers and ultracompact H$\rm _{II}$ regions (see O'Dell
2001 for a recent review). 

The OMC-1S region, located $\sim 100''$ south
of the BN/KL complex, is a young and highly active star formation region.
This region includes a number of deeply embedded mid-IR sources
(Smith et al. 2004), a luminous far-IR/submillimeter source FIR 4 
(Mezger et al. 1990), a dense molecular condensation CS 3 (Mundy et al. 1986) 
, at least four molecular outflows 
(Ziurys et al. 1990; Rodr$\rm{\acute{\i}}$guez-Franco et al. 1999;
Zapata et al. 2005, 2006).
At least seven optical 
Herbig-Haro objects (HH 202, HH269, HH529, HH203/204, HH530, HH625, and HH 528) 
originate from this region (Bally et al. 2000; O'Dell \& Doi 2003).
O'Dell \& Doi (2003) suggested that all the above optical Herbig-Haro objects 
(except HH 625 and HH 530) originated from a region only arcseconds across 
in the OMC-1S region, referred to as Optical Outflow Source (OOS).
However, no IR or radio sources are found in the OOS region.  
Therefore the exciting sources of these Herbig-Haro objects remain unclear.  
Thus we conducted polarimetry in the $J$, $H$, \& $Ks$ bands since infrared polarization studies 
could indicate the illuminating/exciting sources of the infrared nebulae and the outflows
(Nagata et al. 1983; Hodapp 1984; Sato et al. 1985; Tamura et al. 1991).

In this paper, we report the
discovery of several infrared reflection nebulae (IRNe). 
We also discuss the exciting source candidates of 
the optical Herbig-Haro objects in the OMC-1S region.

\section{Observations \& Results}\label{obs}
Near infrared (NIR) polarization images were obtained on 2005 December 26 with the NIR camera 
SIRIUS, a Simultaneous three-color InfraRed Imager for Unbiased Survey
(Nagayama et al. 2003) and its polarimeter mounted on the IRSF, 
1.4 m telescope at 
the South African Astronomical Observatory in Sutherland, South Africa.
The image scale of the array is 0$\farcs$45 pixel$^{-1}$, giving a
field of view of $7'.7 \times 7'.7$. The polarimeter is composed of  
an achromatic (1 - 2.5 $\mu$m) waveplate rotator and a polarizer unit which are located 
upstream of the camera.
More details are described elsewhere (Kandori et al. 2006). 
The polarizations were measured by stepping the half waveplate to
four angular positions (0$^{\circ}$, 45$^{\circ}$, 22.5$^{\circ}$ and  
67.5$^{\circ}$). 10 dithered frames were observed per waveplate position, 
and we observed 9 sets for the object, giving
$9 \times 10$ frames of 10 s integration per waveplate position.
Seeing conditions were $\sim$ 1$\farcs$5 (FWHM) in the $J$ band. 

The data were reduced in the standard manner of infrared image reduction:
subtracting a dark-frame and dividing by a flat-frame.
In addition, the data for each waveplate position 
($I_{0}, \ I_{45}, \ I_{22.5} \ \rm{and} \ I_{67.5}$)
were registered, and then combined. 
Stokes $I$, $Q$, $U$ parameters, degree of polarization ($P$), 
and position angle ($\theta$) are calculated as follows (see e.g., Tamura et al. 2003).  
\begin{eqnarray*}
   I\,&=&\, \{I_{0} + I_{45} + I_{22.5} + I_{67.5}\} \  / \ 2,\\
   Q\,&=&\,I_{0} - I_{45}, \ U\,=\,I_{22.5} - I_{67.5},\\
   P\,&=&\,\sqrt{\left(\frac{Q}{I}\right)^{2} + \left(\frac{U}{I}\right)^{2}},
   \ \ {\rm and} \ \theta \,=\, \frac{1}{2}\,{\rm arctan} \left(\frac{U}{Q}\right).
\end{eqnarray*}
For the region of $26'' \times 20''$ of OMC-1S, 
the polarized intensity images, and polarization vector images in the $J$, $H$, \& $Ks$ bands
obtained with the above calculations are shown in figure \ref{image}(B) and 
figures \ref{vector}(A), (B), and (C), respectively. The entire field ($7'.7 \times 7'.7$) of the data 
is discussed elsewhere (Tamura et al. 2006).

\section{Discussion}
\subsection{Illuminating sources of the reflection nebulae}\label{discuss}
\begin{figure}
  \twocolumn[
    \begin{minipage}{0.5\hsize}
    \begin{center}
      \FigureFile(80mm,50mm){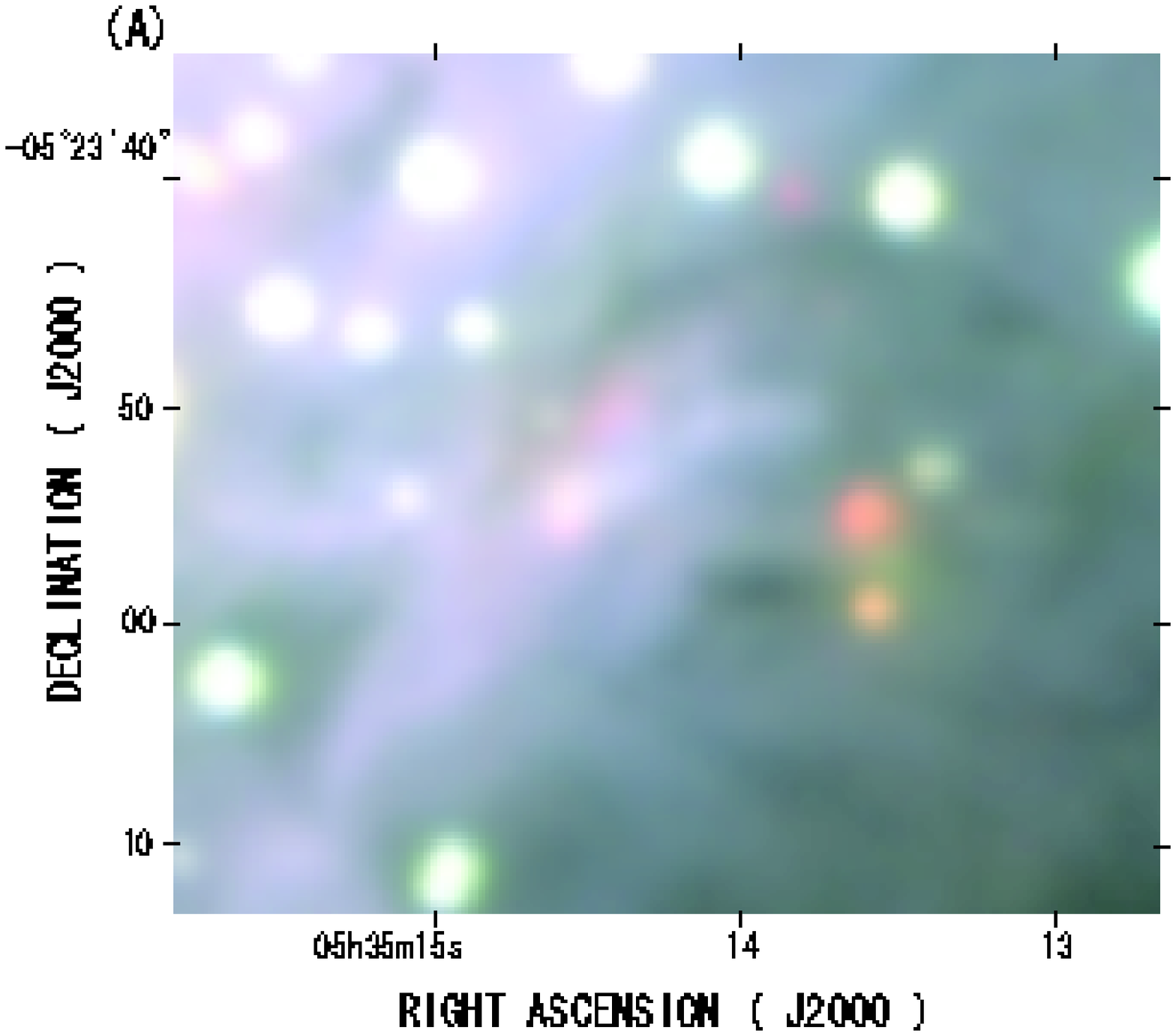}
    \end{center}
    \end{minipage}
    \begin{minipage}{0.5\hsize}
      \begin{center}
	\FigureFile(80mm,50mm){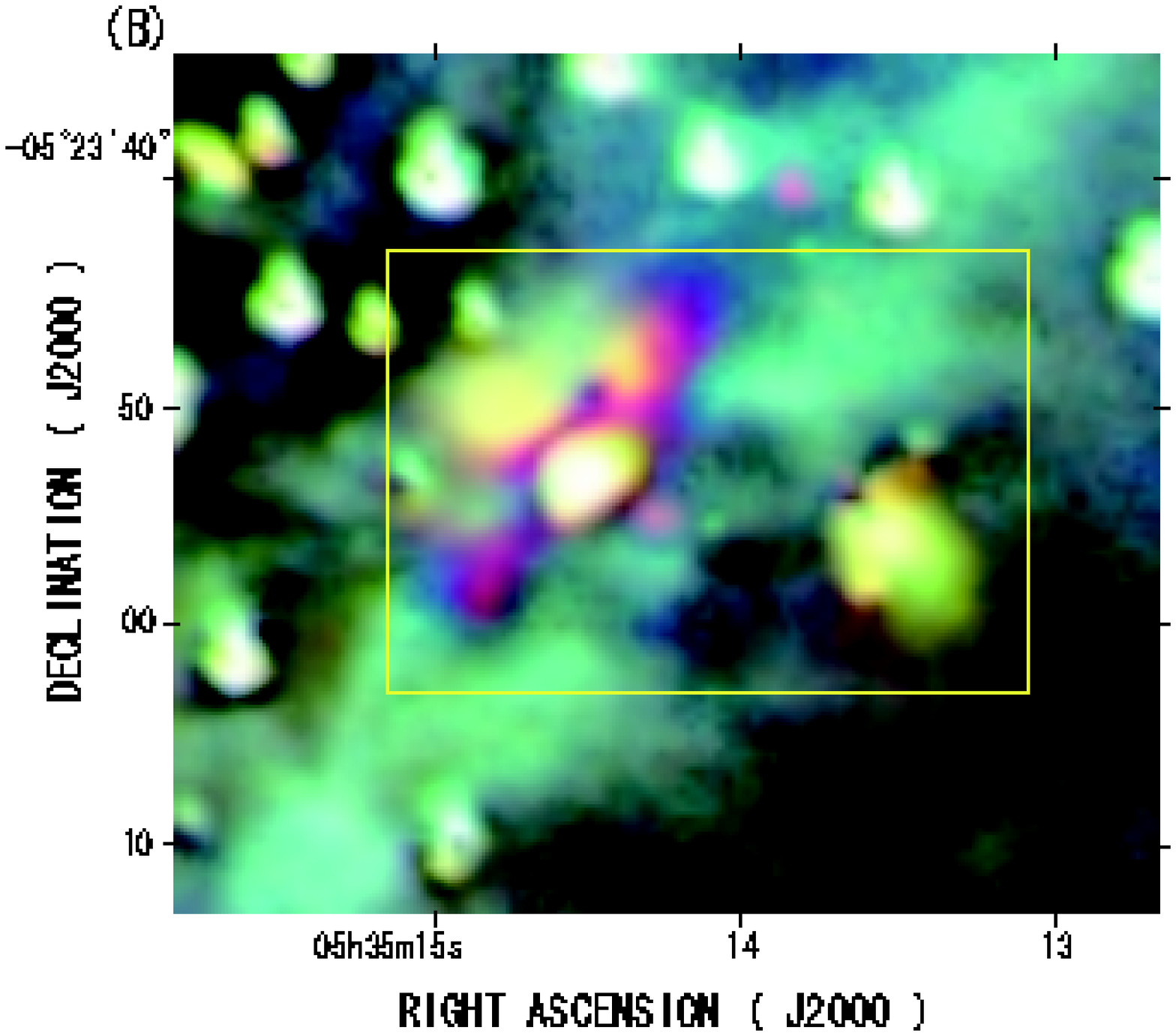}
      \end{center}
    \end{minipage}
    \caption{Near-infrared three-color composite images of the Orion Molecular
      Cloud 1 South (OMC-1S) region in the $J$, $H$, \& $Ks$ bands ($J$: blue, $H$: green, $Ks$: red).
      (A) and (B) show total intensity and polarized 
      intensity images, respectively.
    }\label{image}
  ]
\end{figure}

Figures \ref{image}(A) and (B) show how 
the total intensity and polarization images differ from each other. 
Both figures \ref{image}(A) and (B) are color-composite images in the $J$, $H$ \& $Ks$ bands
($J$: blue, $H$: green, $Ks$: red) which are constructed with the calculations in $\S 2$.
In figure \ref{image}(B), there are several nebulae in the yellow box region,
which are much more clearly seen than in figure \ref{image}(A).
This is because they are obscured by a diffuse nebula
in the intensity image. We identify them as IRNe and 
refer to them as IRN 1 to 6 as shown in figure \ref{hh}. 

Figures \ref{vector}(A), (B), and (C) show polarization vector images superposed on 
the intensity image in the $J$, $H$, \& $Ks$ bands in a yellow box region in figure \ref{image}(B), respectively.
In figures \ref{vector}(A), (B), and (C), 
the polarization vectors show different centrosymmetric patterns 
suggesting that IRN 1 \& 2, IRN 3, 4, \& 5, and IRN 6 are illuminated by 
three independent sources. In order to identify these sources,
we plotted the NIR (3.6 $\mu$m), MIR (8.8, 11.7, and 18.75 $\mu$m) and 
radio (CO 2 $\rightarrow$ 1) sources in figure \ref{vector}(D).
We refer to these sources as marked in figure \ref{vector}(D), e.g. Source 144-351.
The NIR, MIR, and radio sources are detected by the {\it Spitzer Space Telescope},
Smith et al. (2004), and  Zapata et al. (2005), respectively. 
In addition, we plotted the vectors in the $Ks$ band rotated 
by 90$^{\circ}$ in figure \ref{vector}(D), since the direction normal to each polarization
vector indicates the source of illumination. 
The convergence of these lines show that Source 144-351, Source 145-356, and Source 136-355 
are most likely to be the sources of illumination for IRN 1 \& 2, IRN 3, 4, \& 5, 
and IRN 6, respectively.
Since diffuse radiation from the Trapezium cluster is not negligible in the $J$ \& $H$ bands 
(Tamura et al. 2006), we only use the vectors  
in the $Ks$ band in order to identify the source of illumination.

We consider that these IRNe correspond 
to the walls of the outflow material, since 
a close relationship has been suggested 
between IRN and CO bipolar outflow (Nagata et al. 1983; Hodapp 1984; Sato et al. 1985; Yamashita et al. 1989).
The relationship is that these outflows are or were powerful enough to open a cavity within their parent
molecular cloud, in a direction that tends to be perpendicular to the optically thick
circumstellar disk, allowing radiation from the star to escape along the polar directions, 
and then be scattered off dust grains associated with a bipolar outflow.  
Indeed, viewed at low spatial resolution, the extension of IRN tends to be consistent with 
that of CO outflow.

Based on this interpretation, a suggested geometry of each outflow and associated IRN is depicted
in figure \ref{vector}(D); the blue and red cones means blueshifted and redshifted outflows, respectively.
IRN 6 associated with Source 136-355  
has a clear mono-pole structure, which means that IRN 6 corresponds to a
blueshifted outflow. The redshifted counterpart IRN is probably obscured and invisible. 
Other examples of such mono-polar IRNe are R Mon (Aspin et al. 1985),
Cep A (Jones et at. 2004), and GL 2591 (Minchin et al. 1991). 
IRN 3 \& IRN 4 have a bipolar structure extending north-west to south-east; 
we suggest that IRN 3 corresponds to a blueshifted outflow
and IRN 4 corresponds to a redshifted outflow. This is because IRN 3 is 
detected in all of the $J$, $H$, \& $Ks$ bands, while IRN 4 is only in the $Ks$ band, 
which suggests a heavier extinction toward IRN 4. 
Similarly, we consider that IRN 1 and IRN 2 have a likely bipolar structure extending 
east to west; we suggest a blueshifted outflow component in IRN 2 and 
a redshifted outflow component probably in IRN 1. This is because  
IRN 1 is redder than IRN 2 in figure \ref{image}(B). In addition, the small region of IRN 5 
seems to be associated with a local high density region.

Table \ref{degree} summarizes the measured maximum degrees of polarization in IRN 1 to 6.
Subtracting the local background polarization, which is determined at ($-$10$''$, 10$''$)
in figures 2,
produced larger degrees of polarization in the IRNe. 
However, this had little effect on the geometry of the centrosymmetric polarization patterns.
This is most likely to be due to the contribution of unpolarized
ionized gas emission in the background diffuse radiation.
Although the apparent polarization of these IRNe is small (a few \%), 
the level of degrees of polarization after the background 
subtraction is comparable to that of other IRNe such as GL 2591
($\sim 10 \%$ in the $K$ band; Minchin et al. 1991). 

\begin{table}[h]
  \caption{Maximum degrees of polarization in the $JHKs$ bands. 
    The values in the parenthesis are after the background subtraction.}
  \label{degree}
 \begin{center}
   \begin{tabular}{cccc} \hline \hline
     IRNe  & $J$ band  \ \ & $H$ band  \ \  & $Ks$ band  \ \ \\ \hline
       1 &     -----                & $ \sim$ 2 \% (17 \%)& $ \sim$ \ 7 \% (27 \%)\\
       2 & $ \sim$  5 \% (10 \%)    & $ \sim$ 3 \% (10 \%)& $ \sim$ \ 7 \% (19 \%)\\ 
       3 &     -----                & $ \sim$ 8 \% (18 \%)& $ \sim$ \ 9 \% (16 \%)\\
       4 &     -----                &        -----          & $ \sim$ \ 3 \% (\ 7 \%)\\
      5 &     -----                &        -----          & $ \sim$ \ 4 \% (12 \%)\\
       6 &     -----                & $ \sim$ 8 \% (19 \%)& $ \sim$ 10 \% (20 \%)\\ \hline
\end{tabular}
  \end{center}
\end{table}

\subsection{The possible exciting sources of the Herbig-Haro objects}
As mentioned in \S1, seven optical 
Herbig-Haro objects (HH 202, HH269, HH529, HH203/204, HH530, HH625, and HH 528)
originate from the OMC-1S region. 
Figure \ref{hh} shows the polarization vector map superposed on  
polarized intensity image in the $Ks$ band,
together with the directions of the HH outflows, and the Optical Outflow Source (OOS) 
suggested by O'Dell \& Doi (2003), marked with an error bar (1$''$.5 $\times$ 1$''$.5).
Here we discuss possible associations between these outflows 
and the embedded sources indicated in \S \ref{discuss}, based on our polarization
data and direction of IRNe. 

\begin{table}
\twocolumn[
  \caption{IR sources, their associated IRNe, and Herbig-Haro objects}
  \label{summary}
  \begin{center}
    \begin{tabular}{cccc} \hline \hline
      IR sources      & detected wavelengths  \ \ \        & IRNe  \ \ \   & HH objects  \ \ \\ \hline
       Source 144-351 & NIR, MIR, Radio & IRN 1, IRN 2        & \bf HH 529                 \\
       Source 145-356 &      MIR, Radio & IRN 3, IRN 4, IRN 5 & HH 202, HH 528 or HH 203/204 \\ 
       Source 136-355 & NIR, MIR, Radio & IRN 6               & \bf HH 530, HH 269        \\
       Source 136-360 & NIR, MIR, Radio & None                & HH 625                       \\ \hline
       
\end{tabular} 
  \end{center}
]
\end{table}

{\sl HH 529.} ---
HH 529 is blueshifted (Walter et al. 1995), and extended to the east
from OMC-1S at PA $\sim 100^{\circ}$.
The direction of HH 529 corresponds approximately to the extension of IRN 2.
Hence we conclude that HH 529 originates from Source 144-351.

{\sl HH 269.} ---
HH 269 is also blueshifted (Walter et al. 1995), and extended in the west
at PA $\sim 280^{\circ}$. 
Recently, Zapata et al. (2006) found a SiO bipolar outflow
extended in the east and west from Source 136-355, 
and concluded that
the blueshifted SiO outflow at PA $\sim 280^{\circ}$  
could be powering HH 269. 
However, we found IRN 6 extended to the southwest from Source 136-355.
There might be an another source powering HH 269 behind Source 136-355 which illuminates IRN 6.

{\sl HH 202.} ---
HH 202 is one of the first HH objects to be recognized in the Orion Nebula
(Canto et al. 1980), and is a highly collimated jet (Rosado et al. 2001), 
extending at PA $\sim 330^{\circ}$. This direction corresponds 
approximately to the extension of IRN 3. So we conclude that 
HH 202 originates from Source 145-356. 

{\sl HH 528 and HH 203/204.} ---
HH 528 has a low-velocity redshifted component (Smith et al. 2004) 
at PA $\sim 160^{\circ}$ and HH 203/204 has a blueshifted component (O'Dell et al. 1997)
at PA $\sim 135^{\circ}$ .
We consider that HH 528 is most likely to be a counterpart to HH 202, since the blueshifted
HH 202 should have a redshifted counterflow.  

{\sl HH 530.} ---
HH 530 extending at PA $\sim 250^{\circ}$ was first identified by Bally et al. (2000),
where they suggested that HH 530 originated from the FIR 4/CS 3 region 
located $\sim 20''$ south of Source 136-355, since the highly collimated, 
low velocity redshifted CO outflow from the FIR 4/CS 3 region
was reported in Schmid-Burgk et al. (1990). However, HH 530 is not aligned in the direction 
of the measured low-velocity outflow. Hence we consider that HH 530 originates from
Source 136-355 since the direction of HH 530 corresponds approximately to  
the extension of IRN 6.

\newpage

\begin{figure}
  \twocolumn[
    \begin{center}
    \FigureFile(140mm,50mm){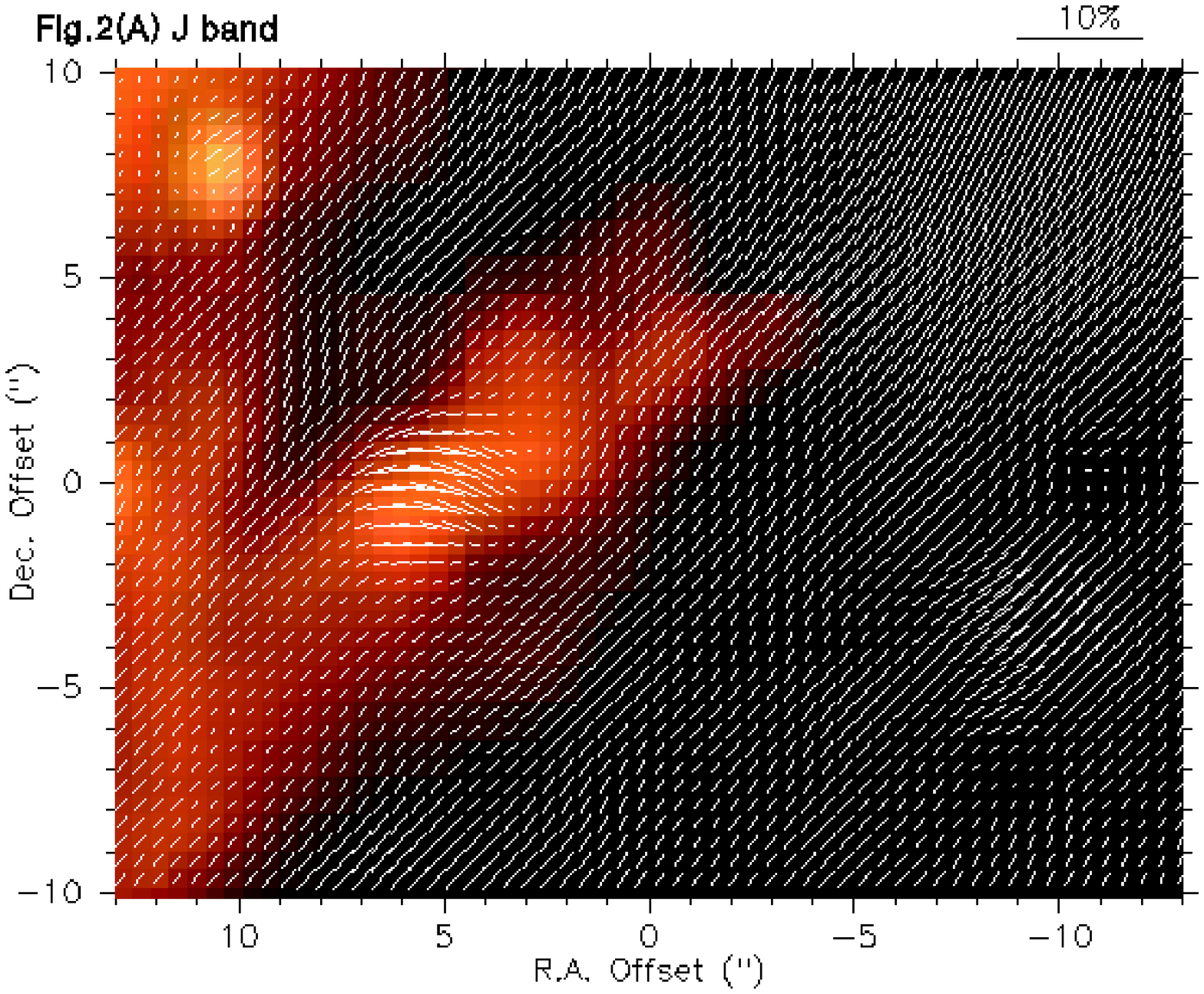}
    \FigureFile(140mm,50mm){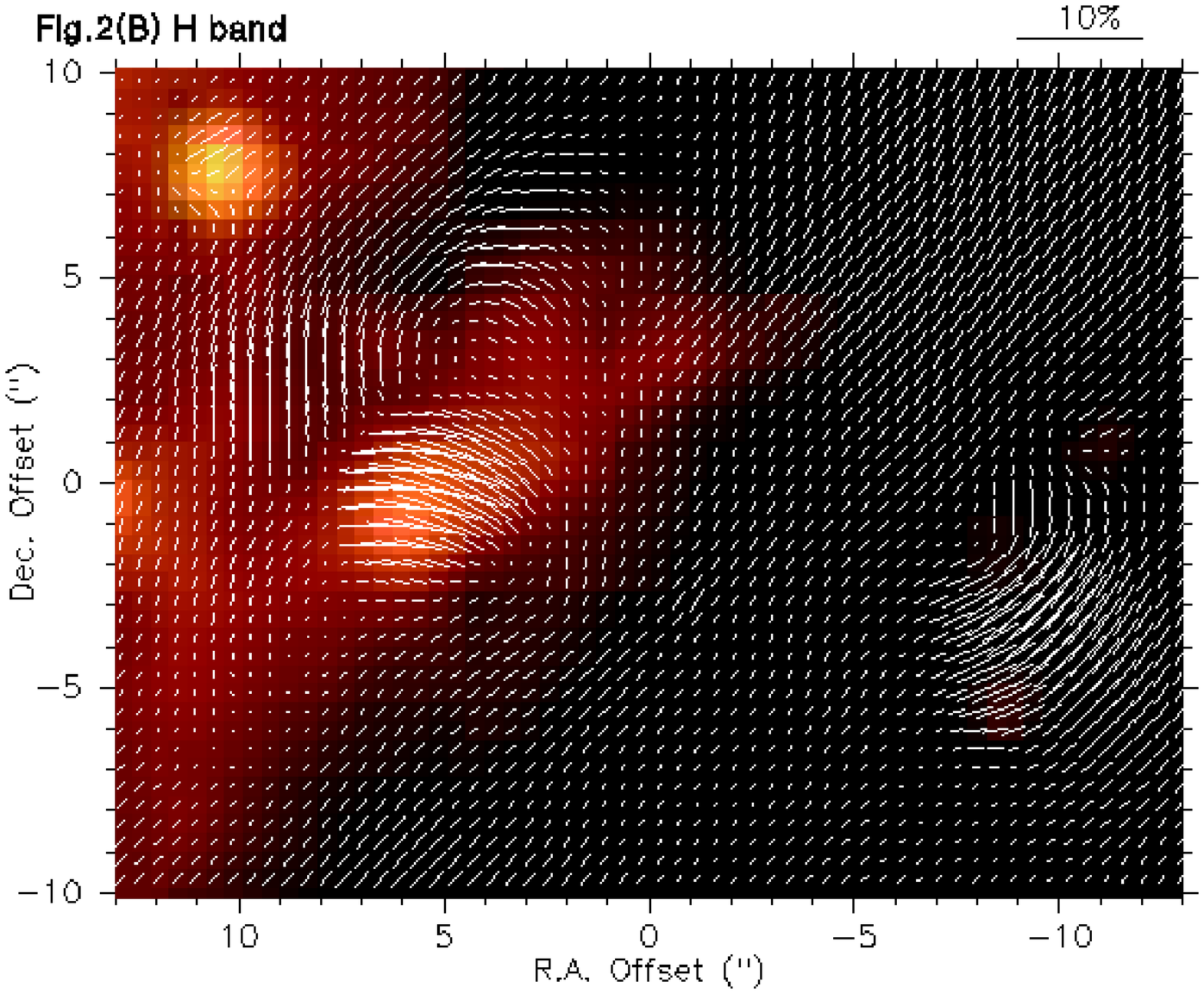}
    \end{center}
]
\end{figure}
\newpage

\

\newpage
\begin{figure}
\twocolumn[
    \begin{center}
    \FigureFile(140mm,50mm){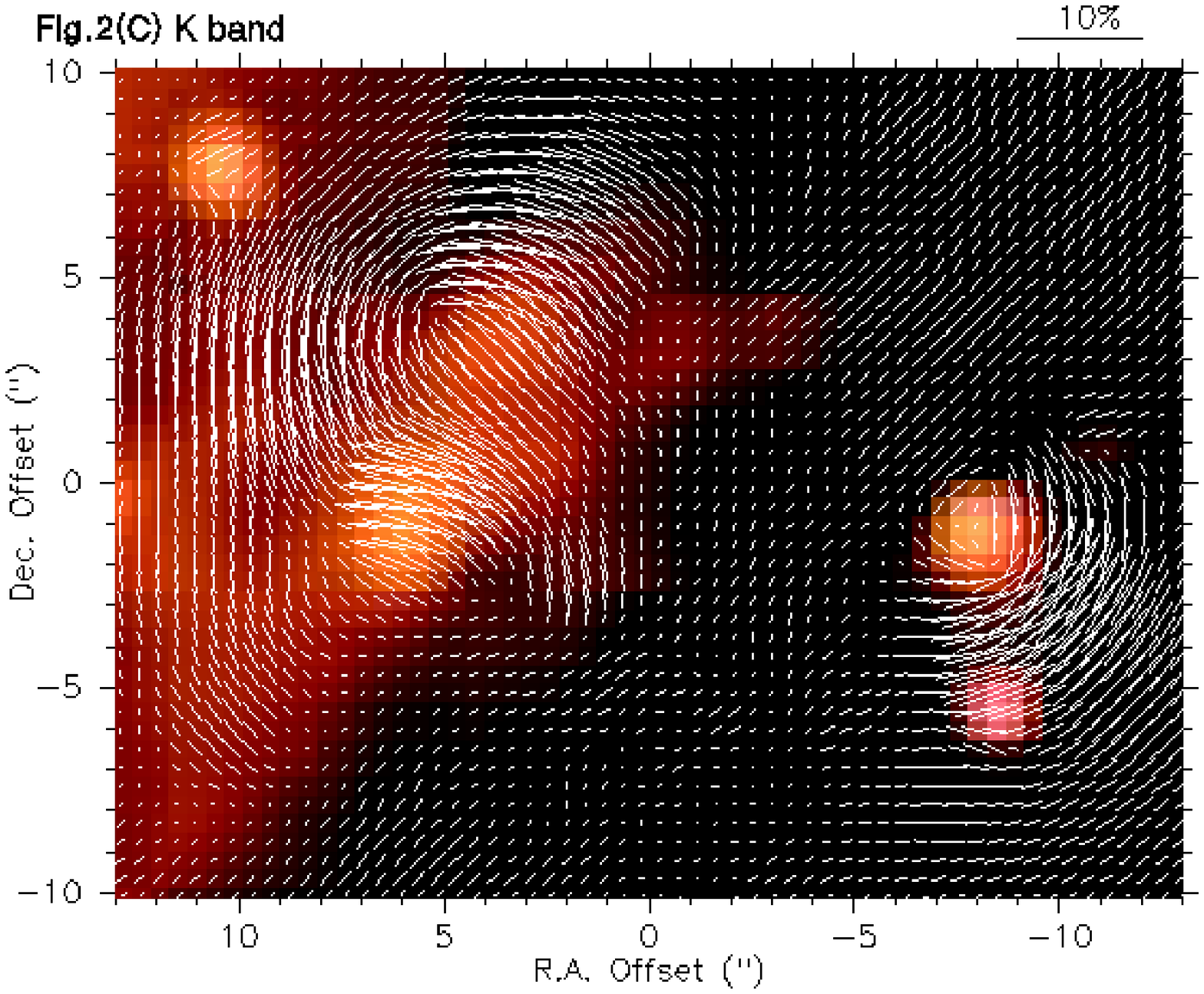}
    \FigureFile(140mm,50mm){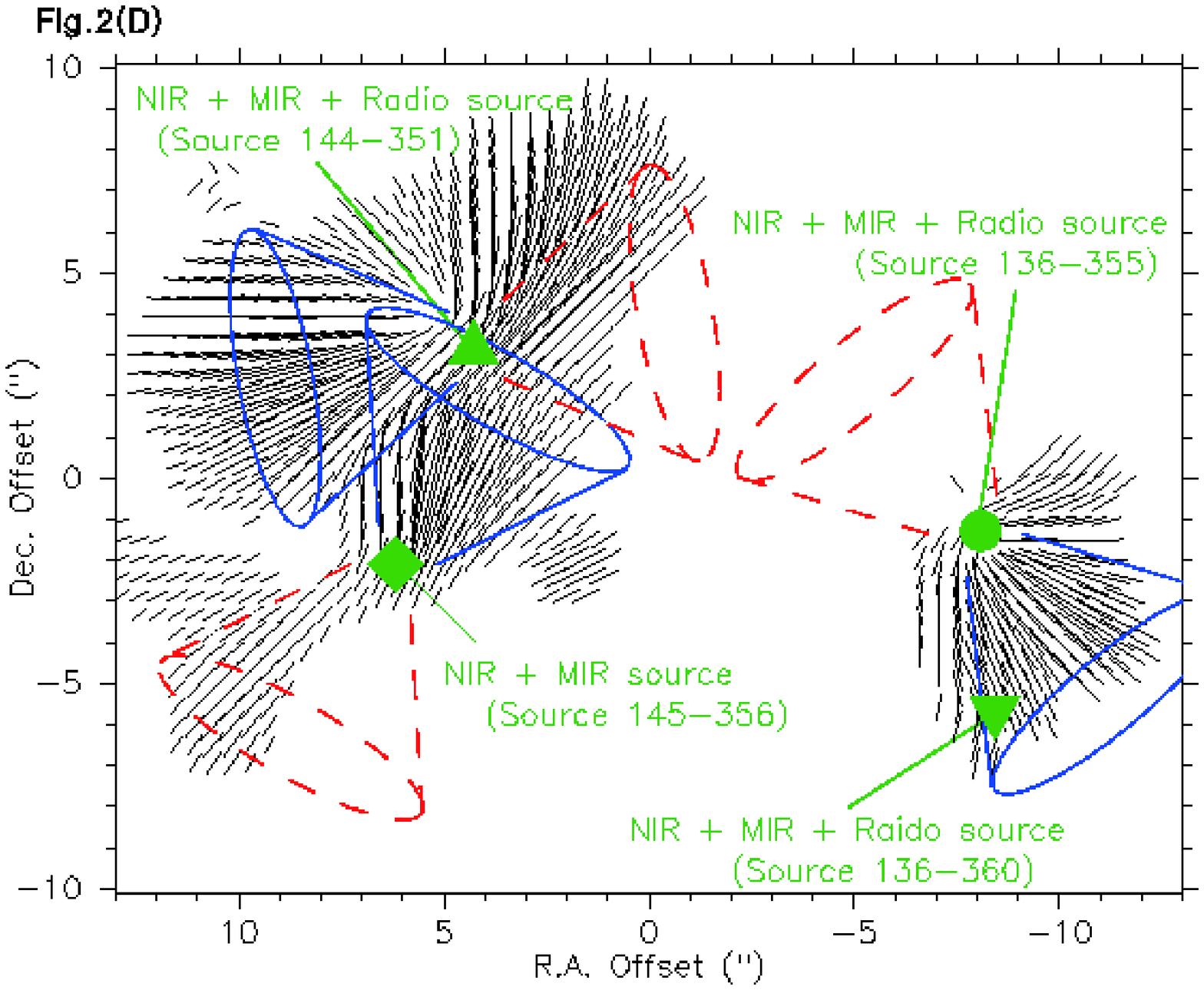}
    \end{center}
    \caption{(A), (B), and (C) : Polarization vectors superposed on
      the intensity image in the $J$, $H$, \& $Ks$ bands, respectively.
      For these polarization vector images, the local background polarizations
      are not subtracted.
      On these scales, $3''$ represent the length of 10 \% polarization vector.
      (D) : The positions of NIR, MIR, and radio sources. The polarization vectors
      in the $Ks$ band are rotated 90$^{\circ}$ in order to identify the illuminating sources of IRNe.
      For constructing this 90 degree rotated vector images, the local
      background, which is determined at ($-10'', 10''$), is subtracted.
      The polarization vectors in the low polarized intensity regions are not plotted for presentation purpose.
      The ($0'', 0''$) coordinate is $5^{\rm h} 35^{\rm m} 14^{\rm s}.1,
      -5^{\circ} 23' 54''.2$ (J2000). }
    \label{vector}
]
\end{figure}

\newpage

\

\newpage

\begin{figure}
\twocolumn[
  \begin{center}
    \FigureFile(170mm,50mm){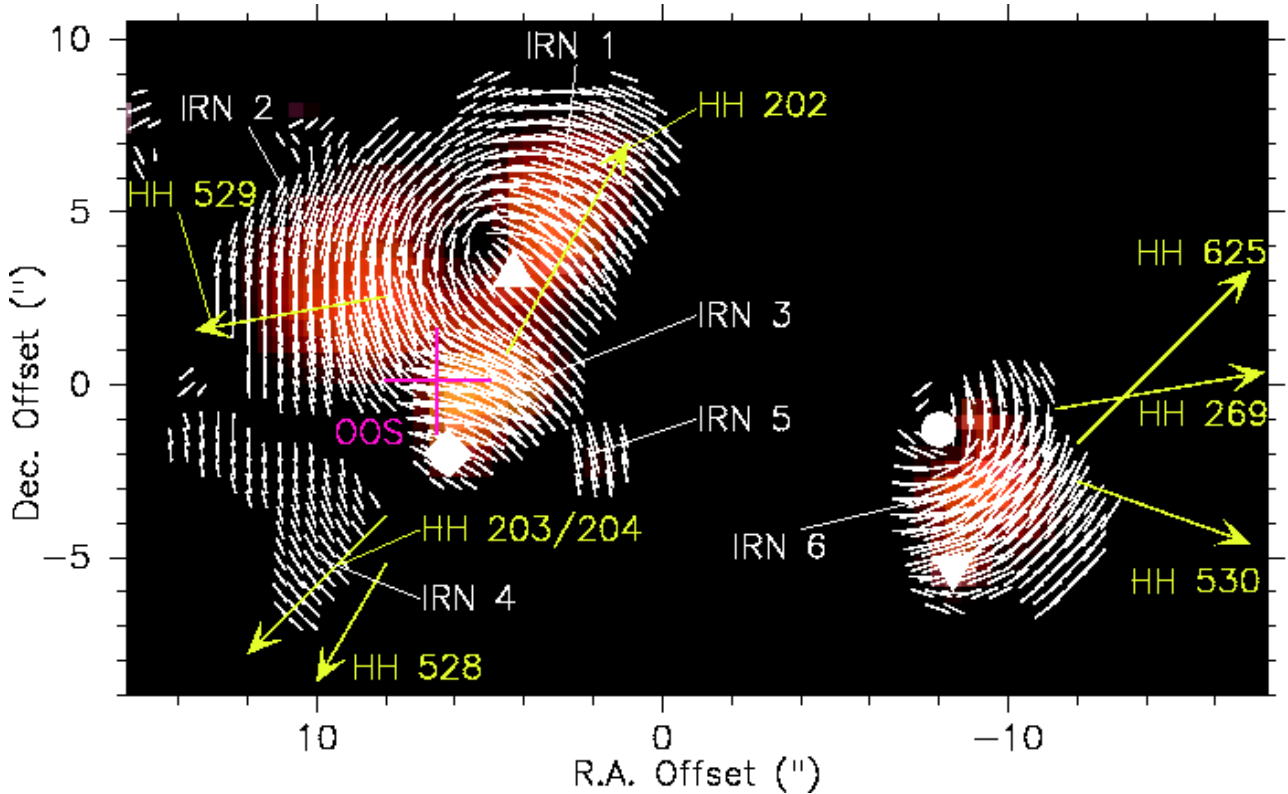}
  \end{center}
  \caption{
    The yellow arrows show the direction of the HH objects originating from
    the OMC-1S region. The polarization vector image superposed 
    on polarized intensity color image in the $Ks$ band.
    For these vector images, local background, which is determined 
    at ($-$10$''$, 10$''$), is subtracted.
    The polarization vectors on the low polarized intensity regions 
    are not plotted for presentation purpose.
    The cross denotes the position of the OOS (O'Dell \& Doi 2003).
    The solid symbols are the same in figure \ref{vector}(D) 
    and the ($0'', 0''$) coordinate is the same in figure \ref{vector}.
  }
  \label{hh}
]
\end{figure}

{\sl HH 625.} ---
HH 625 is a blueshifted HH object extending at PA $\sim 325^{\circ}$ (O'Dell \& Doi 2003).
Zapata et al. (2005) conducted high resolution ($\sim 1''$)
CO line observations, and reported a blueshifted CO outflow emanated from
Source 136-360 at PA $\sim 305^{\circ}$. Hence Source 136-360 is a likely
source of HH 625. Unfortunately, we were unable to detected any IRNe associated with 
Source 136-360.  

\section{Conclusions}
We have conducted the polarimetric imaging of OMC-1S.  
We found that there are at least six IRNe in this region and identified the
illuminating sources of these nebulae; IRN 1 and 2, IRN 3 to 5, and IRN 6 are illuminated by
the IR sources, Source 144-351, 145-356, and 136-355, respectively.
In addition, from a comparison of the extension of IRNe and the direction
of Herbig-Haro objects, we suggest the exciting source of the optical Herbig-Haro objects; 
HH529, a pair of HH202 and HH528 or HH 203/204, HH 530 and HH269 are 
originated from Source 144-351, 145-356, and 136-355, respectively. 
Table \ref{summary} summarizes the illuminating sources of IRNe and the driving sources of Herbig-Haro objects. 
Further high spatial resolution CO observations 
should be undertaken for revealing the kinematics of outflows in OMC-1S.

\bigskip

We are very grateful to the referee, Jhon Bally, for helpful suggestions. We also thank Robert O'Dell
for useful comments. This work was partly supported by MEXT japan,  Grant-in-Aid for Scientific Research
on Priority Areas, and No. 16340061.

\end{document}